\documentclass[12pt]{article}
\usepackage[english]{babel}
\usepackage{amsmath}
\usepackage{amssymb}
\usepackage{bbm}
\usepackage{color}
\usepackage{graphicx}
\usepackage{epsfig}
\usepackage[titletoc,toc,title]{appendix}
\Roman{section}
\newcommand{\eq}[1]{\begin{equation}#1\end{equation}}
\newcommand{\naw}[1]{\left(#1\right)}
\newcommand{\ket}[1]{\left|#1\right>}
\newcommand{\bra}[1]{\left<#1\right|}
\newcommand{\av}[1]{\left<#1\right>}
\newcommand{\com}[1]{\left[#1\right]}
\newcommand{\modu}[1]{\left|#1\right|}
\newcommand{\poisson}[1]{\left\{#1\right\}}

\begin{document}

\begin{center}
\textsc{\Large{Violation of Bell inequality based on $S_4$ symmetry}}
\newline

\large{Katarzyna Bolonek-Laso\'n}\footnote{kbolonek@uni.lodz.pl}\\ 
\emph{\normalsize{Faculty of Economics and Sociology, Department of Statistical Methods, \\University of Lodz,
41/43 Rewolucji 1905 St., 90-214 Lodz,  Poland.}}\\

\end{center}
\begin{abstract}
In two recent papers (\emph{Phys. Rev.} \textbf{A90} (2014), 062121 and \emph{Phys. Rev.} \textbf{A91} (2015), 052110) an interesting method of analyzing the violation of Bell inequalities has been proposed which is based on the theory of finite group representations. We apply here this method to more complicated example of $S_4$ symmetry. We show how the Bell inequality can be related to the symmetries of regular tetrahedron. By choosing the orbits of threedimensional representation of $S_4$ determined by the geometry of tetrahedron we find that the Bell inequality under consideration is violated in quantum theory. The corresponding nonlocal game is analyzed.
\end{abstract} 

\section{Introduction}
In the seminal paper \cite{bell} Bell showed that any local realistic theory must satisfy certain conditions known as Bell inequalities. Since then numerous Bell inequalities have been derived which are characterized by the number of parties, measurement settings and outcomes for each measurement \cite{clauser}$\div$\cite{cabello} (for a review, see \cite{liang} and \cite{brunner}).

The importance of Bell inequalities relies on the observation that they can be violated in the quantum mechanical case. Due to this property they can be used for the tests of entanglement and as a basis for protocols in quantum cryptography \cite{ekert}.

Important contribution to the subject has been made by Fine \cite{fine}, \cite{fine1} (see also \cite{halliwell}, \cite{halliwell1}). Basically, his main conclusion was that, given a number of random variables possessing the joint probability distribution, the Bell inequalities can be derived from the assumption that the relevant probabilities (entering the Bell inequalities) are obtained as the marginals from joint probability distribution. What is more important, the Bell inequalities provide also the sufficient condition for the existence of joint probability distribution returning the initial probabilities as marginals. This is crucial observation for understanding the violation of Bell inequalities in quantum mechanics. Here the joint probabilities can be constructed only for the sets of mutually commuting observables. Therefore, no Bell-type inequality could be derived for joint probabilities of commuting variables if they would follow from the assumption that the relevant probabilities emerge as the marginals of joint distribution for larger set of generally noncommuting variables. 

The general scheme for deriving the Bell inequalities is now quite simple. The relevant combination of probabilities is written in terms of marginals of the joint probability distribution (assumed to exist) arriving at the expression of the form $\sum\limits_{\alpha\in\underline{\alpha}}c(\alpha)p(\alpha)$, where $c(\alpha)$ are integers indicating the number of times $p(\alpha)$ appears in the sum. Due to $0\leq p(\alpha)\leq 1$, $\sum\limits_{\alpha}p(\alpha)=1$ one obtains 
\eq{\min\limits_{\alpha}c(\alpha)\leq\sum_\alpha c(\alpha)p(\alpha)\leq\max\limits_{\alpha}c(\alpha)\label{ad1}}
which is the Bell inequality. In order to get the standard form one should express the probabilities $p(\alpha)$ in terms of relevant correlation functions. \\
For example, the celebrated CHSH inequality \cite{clauser}
\begin{equation}
\modu{E\naw{a,b}+E\naw{a',b}+E\naw{a,b'}-E\naw{a',b'}}\leq 2
\end{equation}	
can be obtained by (i) expressing the correlation functions $E(a,b)$ etc. by the probabilities $p(a,b)$ (ii) writing the probabilities $p(a,b)$ etc. as the marginals of the joint probability $p(a,a',b,b')$. Note that, on the quantum level, all $p(a,b)$, $p(a',b)$, $p(a,b')$ and $p(a',b')$ exist because they correspond to commuting variables while $p(a,a',b,b')$ makes no sense. 

Once the way of deriving the Bell inequality for a given combination of probabilities is established one should look for the examples of its quantum mechanical violation. Recently, in two nice papers \cite{ugur}, \cite{ugur1} G\H uney and Hillery proposed a method based on group theory. One takes some finite group $G$ and selects its irreducible representation $D$.
The results (the form of the left hand sides of Bell inequalities, the degree of their violation etc.) will depend on the choice of the group $G$. Given a group $G$ one still deals with a number of its irreducible representations. Each representation will, in general, give rise to a different definition of the relevant observables and probabilities entering Bell inequalities; the choice of the representation has an impact on the dimension of quantum state and the number of the outcomes of measurements. For example, the group $S_4$ we are here interested in, possesses, as explained below, the twodimensional irreducible representation. It is, however, a homomorphic representation, its kernel being of order four. In fact, it is the faithful representation of the $S_3$ group which is, in turn, isomorphic to the dihedral group $D_3$. If we had chosen this representation we would have obtained the results described (among others) in Ref. \cite{ugur1}. In the threedimensional representation we are going to consider $S_4$ has a transparent geometric interpretation (as a symmetry of the regular tetrahedron) which helps greatly to construct the orbits with desired properties (described below).

 The space carrying the representation $D$ becomes the space of states of one party. The total space of states describing both parties carries the reducible representation obtained by multiplying $D$ by itself. Let $\ket{\varphi}\otimes\ket{\psi}$ be any product state. One constructs the operator 
\begin{equation}
X\naw{\varphi,\psi}\equiv \sum_{g\in G}\naw{D(g)\ket{\varphi}\otimes D(g)\ket{\psi}}\naw{\bra{\varphi}D^\dag(g)\otimes\bra{\psi}D^\dag(g)}\label{c}
\end{equation}
Defining the orbits
\begin{equation}
\ket{g,\varphi}\equiv D(g)\ket{\varphi},\qquad \ket{g,\psi}\equiv D(g)\ket{\psi}
\end{equation}
\begin{equation}
\ket{g,\varphi,\psi}\equiv\ket{g,\varphi}\otimes\ket{g,\psi}
\end{equation}
one finds for an arbitrary bipartite state $\ket{\chi}$:
\begin{equation}
\bra{\chi}X\ket{\chi}=\sum_{g\in G}\modu{\av{g,\varphi,\psi|\chi}}^2\label{a}
\end{equation} 
The sum of probabilities on the right-hand side of eq. (\ref{a}) can be easily maximized. To this end one has to find the maximal eigenvalue of $X\naw{\varphi,\psi}$. Note that $X\naw{\varphi,\psi}$ commutes with all $D(g)\otimes D(g)$, $g\in G$. Assume that in the Clebsh-Gordon decomposition into irreducible pieces 
\begin{equation}
D\otimes D=\bigoplus\limits_{s}D^{(s)}\label{f}
\end{equation}
each $D^{(s)}$ appears only once. Then, by Schur lemma, $X\naw{\varphi,\psi}$ is automatically diagonal and reduces to a multiple of unity on each irreducible component. Using orthogonality relations it is easy to see that the relevant eigenvalues of $X\naw{\varphi,\psi}$ are
\begin{equation}
\frac{\modu{G}}{d_s}\|\ket{\varphi}\otimes\ket{\psi}_s\|^2 \label{a3}
\end{equation}
where $\modu{G}$ is the order of $G$, $d_s$ is the dimension of $D^{(s)}$ representation and $\ket{\varphi}\otimes\ket{\psi}_s $ is the projection of 
$\ket{\varphi}\otimes\ket{\psi} $ on to the carrier space of $D^{(s)}$. In general, one can consider a number of vectors $\ket{\varphi_n}\otimes\ket{\psi_n}$ and the corresponding operators $X\naw{\varphi_n,\psi_n}$. They mutually commute so the eigenvalues of 
\begin{equation}
X=\sum_{n=1}^N X\naw{\varphi_n,\psi_n}\label{b4}
\end{equation}
are simply the sums of the eigenvalues of all $X\naw{\varphi_n,\psi_n}$. In this way one can maximize the sum of probabilities 
\begin{equation}
\sum_{n=1}^N\sum_{g\in G}\modu{\av{g,\varphi_n,\psi_n|\chi}}^2.\label{b}
\end{equation}
Particularly interesting situation arises if $\ket{\varphi_n}$ and $\ket{\psi_n}$ are chosen in such a way that the orbits $\poisson{\ket{g,\varphi_n}}_{g\in G}$ and $\poisson{\ket{g,\psi_n}}_{g\in G}$ are decomposed into disjoint sets of mutually orthogonal vectors, each spanning the carrier space of $D$. Any such set can be viewed as defining some observable via its eigenvectors. In such a case the sum  (\ref{b}) acquires a particularly simple form (cf. Ref. \cite{ugur1}).

Let us describe this in more detail. Once the initial vector $\ket{\varphi}$ (we omit for simplicity the index $n$ numbering the orbits) is selected the size of the orbit $\poisson{\ket{g,\varphi}}$ is determined by the stabilizer subgroup $G_s\subset G$ defined by the condition $D(g)\ket{\varphi}=\ket{\varphi}$; the number of distinct elements of the orbit equals simply $\frac{\modu{G}}{\modu{G_s}}$. Assume, as indicated above, that the orbit is selected in such a way that it is a collection of disjoint sets of vectors, each set forming an orthonormal basis in the space carrying the representation $D$. Therefore, one can write $\poisson{\ket{g,\varphi}}_{g\in G}=\bigcup\limits_{i=1}^{M}\naw{\poisson{\ket{i,\alpha}}_{\alpha =1}^d}$ where $d$ is the dimension of the representation $D$ and $M=\frac{\modu{G}}{d\cdot\modu{G_s}}$.\\
By assumption
\begin{equation}
\av{i,\alpha | i,\beta} =\delta_{\alpha\beta};
\end{equation}
however, in general,
\begin{equation}
\av{i,\alpha |j,\beta}\neq 0\quad \text{for}\quad i\neq j.\label{da}
\end{equation}
Since each set $\poisson{\ket{i,\alpha}}_{\alpha=1}^d$ forms an orthonormal basis one can define the corresponding observables via their spectral decomposition
\begin{equation}
a_i\equiv \sum_{\alpha =1}^d a_i(\alpha)\ket{i,\alpha}\bra{i,\alpha}
\end{equation}
Note that $a_i(\alpha)$ can be chosen to be the arbitrary real numbers. In fact, the Bell inequalities concern, in the last instance, the probabilities and their explicit form depends on the choice of $a_i(\alpha)$ only if we express them in terms of correlation functions instead of probabilities themselves.

Due to the condition (\ref{da}) the observables $a_i$ in general do not commute. As a result it makes no sense to speak on the quantum level about their joint probability distribution. On the other hand, by repeating the above reasoning for the orbit $\poisson{\ket{g,\psi}}_{g\in G}$ one defines the relevant observables $b_i$ for the second party; $b_i$'s and $a_j$'s commute because they act in different components of the tensor product space so the question of their joint probability makes sense even on quantum level. Eq. (\ref{b}) represents a sum of such probabilities for the system in the state described by $\ket{\chi}$. Maximizing this expression can be viewed as looking for Cirel'son bound \cite{cirelson} for the specific set of observables defined by the group theoretical methods. \\
Once the sum (\ref{b}) is properly interpreted one can derive the relevant Bell inequality. Following Fine ideas the existence of joint probability distribution for all observables $a_1,a_2,\ldots ,a_m,b_1,b_1,\ldots,b_m$ is assumed from which the probabilities entering (\ref{b}) are obtained as the marginals and eq. (\ref{ad1}) is applied. On the other hand, by judicious choice of $\varphi$ and $\psi$ one finds that the maximal value of the sum (\ref{b}) violates the Bell inequality derived from eq. (\ref{ad1}). In general, to this end one has to use the observables constructed with the help of more than one orbit for each party $(N>1)$.

Among the examples presented in Refs. \cite{ugur} and \cite{ugur1} a particularly nice is the one based on the dihedral group $D_3$ which is isomorphic to the symmetric group $S_3$. The underlying symmetry of the relevant Bell inequalities is that of regular triangle. In the present paper we consider the Bell inequality based on symmetry of regular tetrahedron - the symmetric group $S_4$.

Let us conclude our introduction by writing out explicitly the Bell inequality we will be dealing with. Each of two parties, Alice and Bob, performs the measurements of eight observables $a_i$, $b_i$, $i=1,\ldots,8$, respectively. Any observable can take three values, 0, 1 or 2. The Bell inequality reads
\begin{small}
\begin{equation}
\begin{split}
& P\naw{a_1=0,b_4=1}+P\naw{a_1=1,b_5=0}+P\naw{a_1=2,b_7=1}
+P\naw{a_2=0,b_4=2}+\\
&  +P\naw{a_2=1,b_8=1}+P\naw{a_2=2,b_5=2}+P\naw{a_3=0,b_4=0}+P\naw{a_3=1,b_8=0}+\\
&  +P\naw{a_3=2,b_7=2}+P\naw{a_4=0,b_3=0}+P\naw{a_4=1,b_1=0}+P\naw{a_4=2,b_2=0}+\\
&  +P\naw{a_5=0,b_1=1}+P\naw{a_5=1,b_6=0}+P\naw{a_5=2,b_2=2}+P\naw{a_6=0,b_5=1}+\\
&  +P\naw{a_6=1,b_7=0}+P\naw{a_6=2,b_8=2}+P\naw{a_7=0,b_6=1}+P\naw{a_7=1,b_1=2}+\\
&  +P\naw{a_7=2,b_3=2}+P\naw{a_8=0,b_3=1}+P\naw{a_8=1,b_2=1}+P\naw{a_8=2,b_6=2}+\\
&  +P\naw{a_1=0,b_8=1}+P\naw{a_1=1,b_8=2}+P\naw{a_1=2,b_8=0}
+P\naw{a_2=0,b_7=2}+\\
&  +P\naw{a_2=1,b_7=0}+P\naw{a_2=2,b_7=1}+P\naw{a_3=0,b_5=0}+P\naw{a_3=1,b_5=2}+\\
&  +P\naw{a_3=2,b_5=1}+P\naw{a_4=0,b_6=2}+P\naw{a_4=1,b_6=1}+P\naw{a_4=2,b_6=0}+\\
& +P\naw{a_5=0,b_3=0}+P\naw{a_5=1,b_3=2}+P\naw{a_5=2,b_3=1}+P\naw{a_6=0,b_4=2}+\\
&  +P\naw{a_6=1,b_4=1}+P\naw{a_6=2,b_4=0}+P\naw{a_7=0,b_2=1}+P\naw{a_7=1,b_2=2}+\\
&  +P\naw{a_7=2,b_2=0}+P\naw{a_8=0,b_1=2}+P\naw{a_8=1,b_1=0}+ P\naw{a_8=2,b_1=1}\leq 14
\end{split}\label{w}
\end{equation}
\end{small}
At the quantum level, for a system exhibiting $S_4$ symmetry, the left hand side can attain the value 14,036.

\section{The $S_4$ group and the tetrahedron symmetry}

The symmetric group $S_4$ is of order 24 and possesses 5 conjugacy classes. There exist five irreducible representations of $S_4$: trivial representation, the alternating representation, the homomorphic twodimensional one and two threedimensional representations, $D$ and $\widetilde{D}$; $\widetilde{D}$ can be obtained from $D$ by multiplication by the alternating representation. All representations can be made real unitary, i.e. orthogonal. 

In what follows we will be interested in the threedimensional representation $D$. Let us write out explicitly the matrices corresponding to transpositions:
\begin{equation}
D\naw{12}=\left[\begin{array}{ccc}
1 & 0 & 0\\
0 & 1 & 0\\
0 & 0 & -1
\end{array}\right],\qquad
 D\naw{13}=\left[\begin{array}{ccc}
1 & 0 & 0\\
0 & -\frac{1}{2} & -\frac{\sqrt{3}}{2}\\
0 & -\frac{\sqrt{3}}{2} & \frac{1}{2}
\end{array}\right]
\end{equation} 
\begin{equation}
D\naw{14}=\left[\begin{array}{ccc}
-\frac{1}{3} & -\frac{\sqrt{2}}{3} & -\frac{\sqrt{6}}{3}\\
-\frac{\sqrt{2}}{3} & \frac{5}{6} & -\frac{\sqrt{3}}{6}\\
-\frac{\sqrt{6}}{3} & -\frac{\sqrt{3}}{6}& \frac{1}{2}
\end{array}\right],\qquad
 D\naw{23}=\left[\begin{array}{ccc}
1 & 0 & 0\\
0 & -\frac{1}{2} & \frac{\sqrt{3}}{2}\\
0 & \frac{\sqrt{3}}{2} & \frac{1}{2}
\end{array}\right]\label{a2}
\end{equation}
\begin{equation}
D\naw{24}=\left[\begin{array}{ccc}
-\frac{1}{3} & -\frac{\sqrt{2}}{3} & \frac{\sqrt{6}}{3}\\
-\frac{\sqrt{2}}{3} & \frac{5}{6} & \frac{\sqrt{3}}{6}\\
\frac{\sqrt{6}}{3} & \frac{\sqrt{3}}{6}& \frac{1}{2}
\end{array}\right],\qquad
 D\naw{34}=\left[\begin{array}{ccc}
-\frac{1}{3} & \frac{\sqrt{8}}{3} & 0\\
\frac{\sqrt{8}}{3} & \frac{1}{3} & 0\\
0 & 0 & 1
\end{array}\right].
\end{equation}
$S_4$ is the symmetry group of regular tetrahedron; it is obvious as any permutation of its vertices leaves it invariant. Any symmetry operation is the composition of reflections in symmetry planes and rotations around symmetry axes. For example, the transpositions correspond to reflections in symmetry planes. One can make the correspondence between the symmetries of tetrahedron and the representation $D$. To this end it is sufficient to show that there exists an orbit which forms a tetrahedron. We start with the vector $\vec{a}_4=\naw{1,0,0}$. It is invariant under $D(12)$, $D(13)$, $D(23)$. The remaining vertices are obtained by the action of $D(14)$, $D(24)$, and $D(34)$. They read: $\vec{a}_1=\naw{-\frac{1}{3},-\frac{\sqrt{2}}{3},-\frac{\sqrt{6}}{3}}$,  $\vec{a}_2=\naw{-\frac{1}{3},-\frac{\sqrt{2}}{3},\frac{\sqrt{6}}{3}}$  and  $\vec{a}_3=\naw{-\frac{1}{3},\frac{\sqrt{8}}{3},0}$ (see Fig.1).
\begin{figure}[htbp]
\centering
\includegraphics[angle=0,width=0.6\textwidth]{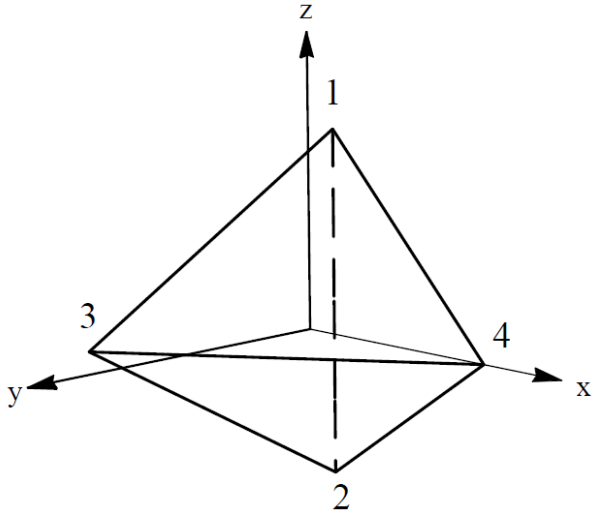}
\caption{The tetrahedron orbit of $S_4$}
\end{figure} 

The tetrahedron forms a nongeneric (degenerate) orbit consisting of four elements (the generic one consists of 24 elements). The reason is that the vector $\vec{a}_4$ we have started with has the stabilizer subgroup consisting of three reflections in the symmetry planes passing through the edges connecting $\vec{a}_4$ with the remaining vertices ($D(12)$, $D(13)$, D(23) mentioned above) and three rotations around the symmetry axis passing through $\vec{a}_4$. Thus we have interpreted the representation of $S_4$ under consideration as the set of symmetry transformations of the tetrahedron. This simplifies greatly the search for the generic orbits with desired properties which are used to define the relevant observables (as sketched in the Introduction). 

According to the prescription given in Ref. \cite{ugur1} we form the operator (cf. eq. (\ref{c}))
\begin{equation}
X\naw{\varphi,\psi}\equiv \sum_{g\in G}\naw{D(g)\ket{\varphi}\otimes D(g)\ket{\psi}}\naw{\bra{\varphi}D^\dag(g)\otimes\bra{\psi}D^\dag(g)}
\end{equation} 
Since all representations of $S_4$ are real we can choose $\ket{\varphi}$ and $\ket{\psi}$ to be real threevectors. The Clebsh-Gordan decomposition for $D\otimes D$ reads
\begin{equation}
D\otimes D=D\oplus \widetilde{D}\oplus D_2\oplus D_0\label{a4}
\end{equation}
where $\widetilde{D}$ is the second threedimensional representation, $D_2$ is the twodimensional representation while $D_0$ is the trivial one. It is not difficult to find the transformation from the product basis to the one exhibiting explicitly the decompostion appearing on the right hand side of eq. (\ref{f}). It is given by
\begin{equation}
C=\left [\begin{array}{ccccccccc}
\sqrt{\frac{2}{3}} & 0 & 0 & 0 & -\frac{1}{\sqrt{6}} & 0 & 0 & 0 & -\frac{1}{\sqrt{6}}\\
0 & -\frac{1}{\sqrt{6}} & 0 & -\frac{1}{\sqrt{6}} & \frac{1}{\sqrt{3}} & 0 & 0 & 0 & -\frac{1}{\sqrt{3}}\\
0 & 0 & -\frac{1}{\sqrt{6}} & 0 & 0 & -\frac{1}{\sqrt{3}} & -\frac{1}{\sqrt{6}} & -\frac{1}{\sqrt{3}} & 0\\
0 & \frac{1}{\sqrt{2}} & 0 & -\frac{1}{\sqrt{2}} & 0 & 0 & 0 & 0 & 0\\
0 & 0 & \frac{1}{\sqrt{2}} & 0 & 0 & 0 & -\frac{1}{\sqrt{2}} & 0 & 0\\
 0 & 0 & 0 & 0 & 0 & \frac{1}{\sqrt{2}} & 0 & -\frac{1}{\sqrt{2}} &  0\\
 0 & \frac{1}{\sqrt{3}} & 0 & \frac{1}{\sqrt{3}} & \frac{1}{\sqrt{6}} & 0 & 0 & 0 & -\frac{1}{\sqrt{6}}\\
 0 & 0 & \frac{1}{\sqrt{3}} & 0 & 0 & -\frac{1}{\sqrt{6}} & \frac{1}{\sqrt{3}} & -\frac{1}{\sqrt{6}} & 0\\
 \frac{1}{\sqrt{3}} & 0 & 0 & 0 & \frac{1}{\sqrt{3}} & 0 & 0 & 0 & \frac{1}{\sqrt{3}}
\end{array}\right]\label{db}
\end{equation}
The eigenvalues of the operator $X\naw{\varphi,\psi}$ are given by the formula (\ref{a3}). Using the above form of $C$ we can compute them explicitly. 
To this end we need only to know the lenght of the projection of the vector $\ket{\varphi}\otimes\ket{\psi}$ on each irreducible component in the decomposition (\ref{a4}). Due to the reality of the representation the vectors $\ket{\varphi}$ and $\ket{\psi}$ can be taken as unit real threedimensional vectors (their explicit form is specified in the next section):
\begin{equation}
\ket{\varphi}=\left(\begin{array}{c}
m_1\\
m_2\\
m_3
\end{array}\right),\qquad \ket{\psi}=\left(\begin{array}{c}
m_1'\\
m_2'\\
m_3'
\end{array}\right), \qquad \vec{m}^2=\vec{m}'^2=1
\end{equation}

Representing then $\ket{\varphi}\otimes\ket{\psi}$ as a one column nine rows matrix and applying the transformation defined by the matrix (\ref{db}) we find that the relevant projections of $\ket{\varphi}\otimes\ket{\psi}$ onto irreducible subspaces read:

\begin{equation}
D_0:\quad \frac{1}{\sqrt{3}}\naw{\vec{m}\cdot\vec{m}'}\hspace{5.5cm}\label{d1}
\end{equation}
\begin{equation}
\begin{split}
& D_2:\quad \frac{1}{\sqrt{3}}\naw{m_1m_3'+m_3m_1'}-\frac{1}{\sqrt{6}}\naw{m_2m_3'+m_3m_2'}\\
& \qquad \quad \frac{1}{\sqrt{3}}\naw{m_1m_2'+m_2m_1'}+\frac{1}{\sqrt{6}}\naw{m_2m_2'-m_3m_3'}
\end{split}\label{b1}
\end{equation}
\begin{equation}
\widetilde{D}:\quad \frac{1}{\sqrt{2}}\naw{\vec{m}\times\vec{m}'}\hspace{5.5cm}
\end{equation}
\begin{equation}
\begin{split}
& D: \quad \sqrt{\frac{2}{3}}m_1m_1'-\frac{1}{\sqrt{6}}\naw{m_2m_2'+m_3m_3'}\\
& \qquad\quad \frac{1}{\sqrt{3}}\naw{m_2m_2'-m_3m_3'}-\frac{1}{\sqrt{6}}\naw{m_1m_2'+m_2m_1'}\\
& \qquad\quad -\frac{1}{\sqrt{3}}\naw{m_2m_3'+m_3m_2'}-\frac{1}{\sqrt{6}}\naw{m_1m_3'+m_3m_1'}.
\end{split}\label{b2}
\end{equation}
The above decomposition can be easily understood. $D$ is the faithful representation of $S_4$ and the latter may be viewed as a subgroup of $O(3)$; the product of two spin one representations of the latter decomposes into the sum of spin zero, one and two ones according to the formula
\begin{equation}
\begin{split}
& m_im_j'=\com{\frac{1}{2}\naw{m_im_j'+m_jm_i'}-\frac{1}{3}\naw{\vec{m}\cdot\vec{m}'}\delta_{ij}}+\\
&\qquad \quad \com{\frac{1}{2}\naw{\vec{m}\times\vec{m}'}_k\epsilon_{kij}}+\com{\frac{1}{3}\naw{\vec{m}\cdot\vec{m}'}\delta_{ij}}\equiv\\
& \qquad \quad \equiv S_{ij}+A_{ij}+\Delta_{ij}.
\end{split}\label{b3}
\end{equation} 
Spin zero $\naw{\Delta_{ij}}$ and one $\naw{A_{ij}}$ representations remain irreducible when subduced to $S_4$ while fivedimensional spin two representation $S_{ij}$ decomposes into two- and threedimensional ones. By comparying eqs. (\ref{b1}) and (\ref{b2}) with eq. (\ref{b3})) we find
\begin{equation}
\begin{split}
& D_2: \quad \frac{2}{\sqrt{3}}S_{13}-\sqrt{\frac{2}{3}}S_{23}\\
& \qquad\quad\frac{2}{\sqrt{3}} S_{12}+\frac{1}{\sqrt{6}}\naw{S_{22}-S_{33}}
\end{split}
\end{equation}   
\begin{equation}
\begin{split}
& D:\quad \sqrt{\frac{3}{2}}S_{33}\\
&\qquad -\sqrt{\frac{2}{3}}S_{12}+\frac{1}{\sqrt{6}}\naw{S_{22}-S_{33}}\\
& \qquad -\sqrt{\frac{2}{3}}S_{13}-\frac{2}{\sqrt{3}}S_{23}. 
\end{split}
\end{equation}
Having $\ket{\varphi}\otimes \ket{\psi}$ decomposed into irreducible pieces one finds from eq. (\ref{a3}) the eigenvalues of $X(\varphi,\psi)$. Once this is done the eigenvalues of general operator $X$ defined by eq. (\ref{b4}) are obtained by simple addition.

The next step consists in selecting the orbits of $S_4$ in the space of states.
According to the reasoning sketched in the Introduction we are looking for the orbits consisting of a number of disjoint subsets, each providing an orthonormal basis and defining the spectral decomposition of some observable. In other words, the states belonging to the orbit are eigenstates of the observables entering the inequalities we are going to derive.

 A generic orbit consists of 24 states. The most optimal choice is the orbit which consists of eight triples of mutually orthogonal vectors, each providing the spectral decomposition of one of eight observables. In order to find such an orbit it is sufficient to find a vector $\ket{x}$ such that $g_1\ket{x}$, $g_2\ket{x}$ and $g_3\ket{x}$ are orthogonal to each other for some $g_1,g_2,g_3\in S_4$. Then for any $g\in S_4$, $gg_1\ket{x}$, $gg_2\ket{x}$ and $g g_3\ket{x}$ are mutually orthogonal. Keeping this in mind we construct the orbit numbered by $x_\alpha^i$, $i=1,\ldots,8$, $\alpha=0,1,2$ such that $\av{x_\alpha^i| x_\beta^i}=\delta_{\alpha\beta}$ (no summation over $i$). We will be dealing with eight observables (for each party) $a_i$, $i=1,\ldots,8$ with spectral decomposition
\begin{equation}
a_i=\sum_{\alpha=0}^2\alpha \ket{x_{\alpha}^i}\bra{x_{\alpha}^i}.\label{b5}
\end{equation}

Now, it is not difficult to find explicitly the vectors $\ket{x_{\alpha}^i}$ by viewing the action of $S_4$ as the symmetry group of the regular tetrahedron, i.e. the set of rotations around symmetry axes and reflections in symmetry planes. In fact, the problem reduces to the Euclidean geometry in three dimensions. It is not dificult to see that starting from the vector (called later $x_2^8$ for convenience) 
\begin{equation}
\ket{x}=\frac{1}{\sqrt{3}}\left [\begin{array}{c}
1\\
1\\
1
\end{array}\right ]\label{ac}
\end{equation}
and reflecting it in the planes perpendicular to the edges $(12)$, $(23)$, $(13)$ one obtains three orthogonal vectors. Therefore, (\ref{ac}) is a good starting point to construct the orbit. In this way we obtain the orbit  $\poisson{\ket{x_{\alpha}^i};\alpha=0,1,2,i=1,\ldots,8}$. The explicit form of the vectors $\ket{x_\alpha^i}$ is given in Appendix A. The states forming this orbit define, via eq. (\ref{b5}), the eight observables $a_i$, $i=1,\ldots,8$.

\section{Construction of a Bell inequality, local bounds and nonlocal correlations}
Following the idea of Refs. \cite{ugur} and \cite{ugur1} we present the example of the violation of Bell inequality based on the quantum states constructed with the help of threedimensional representation $D$ of the symmetric group $S_4$. To this end we select two orbits in the space carrying reducible representation $D\otimes D$ of $S_4$. Each orbit has the form $\ket{i,\alpha;i',\alpha '}=\ket{x_\alpha^i}\otimes \ket{x_{\alpha'}^{i'}}$, $i,i'=1,\ldots,8$, $\alpha,\alpha'=0,1,2$. The observables related to the first factor (Alice) are $a_1,...,a_8$ while those related to the second one (Bob) and also defined by eq. (\ref{b5}) will be denoted by $b_1,\ldots,b_8$.

Specifically, the two orbits are defined as follows:\\
The first orbit $O_1$:
\begin{equation}
\begin{split}
& \ket{1,0;8,1},\quad \ket{1,1;8,2},\quad \ket{1,2;8,0},\quad \ket{2,0;7,2},\quad \ket{2,1;7,0}, \quad \ket{2,2;7,1}\\
& \ket{3,0;5,0},\quad \ket{3,1;5,2},\quad \ket{3,2;5,1},\quad  \ket{4,0;6,2},\quad \ket{4,1;6,1},\quad \ket{4,2;6,0}\\
& \ket{5,0;3,0},\quad \ket{5,1;3,2},\quad \ket{5,2;3,1},\quad \ket{6,0;4,2},\quad \ket{6,1;4,1},\quad \ket{6,2;4,0}\\
& \ket{7,0;2,1},\quad \ket{7,1;2,2},\quad \ket{7,2;2,0},\quad \ket{8,0;1,2},\quad \ket{8,1;1,0},\quad \ket{8,2;1,1}\\
\end{split}\label{c1}
\end{equation}
The second orbit $O_2$:
 \begin{equation}
\begin{split}
& \ket{1,0;4,1},\quad \ket{1,1;5,0},\quad \ket{1,2;7,1},\quad \ket{2,0;4,2},\quad \ket{2,1;8,1}, \quad \ket{2,2;5,2}\\
& \ket{3,0;4,0},\quad \ket{3,1;8,0},\quad \ket{3,2;7,2},\quad \ket{4,0;3,0},\quad \ket{4,1;1,0},\quad \ket{4,2;2,0}\\
& \ket{5,0;1,1},\quad \ket{5,1;6,0},\quad \ket{5,2;2,2},\quad \ket{6,0;5,1},\quad \ket{6,1;7,0},\quad \ket{6,2;8,2}\\
& \ket{7,0;6,1},\quad \ket{7,1;1,2},\quad \ket{7,2;3,2},\quad \ket{8,0;3,1},\quad \ket{8,1;2,1},\quad \ket{8,2;6,2}\\
\end{split}\label{c2}
\end{equation}
Due to $\com{a_i,b_j}=0$ the joint probabilities for $a_i$ and $b_j$ make sense both in quantum and classical physics. We consider the sum of probabilities corresponding to the states of both orbits:
\begin{equation}
S=\sum_{\naw{i,\alpha;i',\alpha'}}p\naw{a_i=\alpha; b_{i'}=\alpha'}\label{c3}
\end{equation} 
where the summation runs over all quadruples $\naw{i,\alpha;i',\alpha'}$ appearing in (\ref{c1}) and (\ref{c2}).

First, we derive the quantum-mechanical upper bound for the sum (\ref{c3}) (this is an analogue of Cirel'son bound \cite{cirelson}   for the particular situation under consideration). It is easy to see \cite{ugur}, \cite{ugur1} that the maximum value of $S$ equals the maximal eigenvalue of the operator (\ref{b4}) with $N=2$ and $\ket{\varphi_k}\otimes\ket{\psi_k}\in O_k$, $k=1,2$, being the elements (arbitrarily chosen) of both orbits. In order to simplify the computation we take $\ket{\varphi_1}\otimes\ket{\psi_1}=\ket{1,1;8,2}$ and $\ket{\varphi_2}\otimes\ket{\psi_2}=\ket{6,2;8,2}$. Then using eqs. (\ref{a3}) and (\ref{d1})$\div$(\ref{b2}) we find that the maximal eigenvalue of 
\begin{equation}
X=\sum_{n=1}^2X\naw{\varphi_n,\psi_n},
\end{equation} 
reads
\begin{equation}
\lambda_{max}=\frac{24}{81}\naw{3+\sqrt{3}}^2+\frac{8}{81}\naw{3+4\sqrt{2}}^2
\end{equation}
or
\begin{equation}
\lambda_{max}\approx 14,036.
\end{equation}
The corresponding eigenvector $\ket{\chi}$ of $X$ belongs to the scalar component in the decomposition of $D\otimes D$ into irreducible components. Note that, due to the eq. (\ref{b3}),
\begin{equation}
\text{Tr}_A\naw{\ket{\chi}\bra{\chi}}=\mathbbm{1}_B,\quad \text{Tr}_{B}\naw{\ket{\chi}\bra{\chi}}=\mathbbm{1}_A 
\end{equation}
i.e. $\ket{\psi}$ is maximally entangled. It is natural to expect that the more entangled state is the more likely is that it violates the relevant Bell inequality. 

We conclude that the maximal value of $S$, as defined in eq. (\ref{c3}) is 
\begin{equation}
S_{max}=\lambda_{max}\approx 14,036.
\end{equation}   
We shall now show that on the classical level the maximal value cannot exceed 14. To this end we note that classically one can ignore the noncommutativity of different $a_i's$ (as well as $b_i's$) and assume the existence of joint probabilities $p\naw{\sigma}\equiv p\naw{a_1=\alpha_1,\ldots,a_8=\alpha_8;b_1=\alpha_1',\ldots, b_8=\alpha_8'}$ from which the probabilities entering the right hand side are obtained as the marginals (cf. Refs. \cite{fine} and \cite{fine1}). Inserting the relevant expessions for the probabilities $p\naw{a_i=\alpha;b_{i'}=\alpha'}$ one obtains
\begin{equation}
S=\sum_{\sigma}c(\sigma)p(\sigma).\label{ab}
\end{equation}
Due to $0\leq p(\sigma)\leq 1$, $\sum_\sigma p(\sigma)=1$ the maximal value of $S$ is equal to the largest coefficient $c(\sigma)$. It is easy to see that the 48 terms appearing on the right hand side of eq. (\ref{c3}) come in 16 triples of mutually excluding cases. Therefore, $c\naw{\sigma}$ cannot exceed 16. We show that, actually, the maximal value of $c(\sigma)$ is 14. To see this let us visualize the pattern in which the individual terms appear on the right-hand side of eq. (\ref{c3}).  To this end we select an arbitrary $x_{\alpha}^i$ appearing as a first factor in some element of the orbit $O_1$. Then we take the second factor in this element and look for the corresponding element of the $O_2$ orbit. Its first factor serves then for the search of appropriate  element in the first orbit $O_1$ and so on. In this way we arrive at the following cycles:
\setlength{\unitlength}{0.8cm}\\
\begin{equation}
\begin{picture}(5,4)
\thicklines
\put(0.2,0.5){\line(0,1){1.2}}
\put(1.5,0.5){\line(0,1){1.2}}
\put(2.8,0.5){\line(0,1){1.2}}
\put(0.3,0.5){\line(1,1){1.1}}
\put(1.6,0.5){\line(1,1){1.1}}
\put(0.3,1.7){\line(2,-1){2.4}}
\put(-2,2){$A:$}
\put(-2,0){$B:$}
\put(0,0){$x_1^8$}
\put(1.2,0){$x_0^7$}
\put(2.4,0){$x_1^4$}
\put(0,2){$x_0^1$}
\put(1.2,2){$x_1^2$}
\put(2.4,2){$x_1^6$}
\end{picture}  
\begin{picture}(5,4)
\thicklines
\put(0.2,0.5){\line(0,1){1.2}}
\put(1.5,0.5){\line(0,1){1.2}}
\put(2.8,0.5){\line(0,1){1.2}}
\put(0.3,0.5){\line(1,1){1.1}}
\put(1.6,0.5){\line(1,1){1.1}}
\put(0.3,1.7){\line(2,-1){2.4}}
\put(0,0){$x_2^8$}
\put(1.2,0){$x_0^4$}
\put(2.4,0){$x_0^5$}
\put(0,2){$x_1^1$}
\put(1.2,2){$x_2^6$}
\put(2.4,2){$x_0^3$}
\end{picture}  
\begin{picture}(5,4)
\thicklines
\put(0.2,0.5){\line(0,1){1.2}}
\put(1.5,0.5){\line(0,1){1.2}}
\put(2.8,0.5){\line(0,1){1.2}}
\put(0.3,0.5){\line(1,1){1.1}}
\put(1.6,0.5){\line(1,1){1.1}}
\put(0.3,1.7){\line(2,-1){2.4}}
\put(0,0){$x_0^8$}
\put(1.2,0){$x_2^5$}
\put(2.4,0){$x_1^7$}
\put(0,2){$x_2^1$}
\put(1.2,2){$x_1^3$}
\put(2.4,2){$x_2^2$}
\end{picture}  
\end{equation}
\begin{equation}
\begin{picture}(5,4)
\thicklines
\put(0.2,0.5){\line(0,1){1.2}}
\put(1.5,0.5){\line(0,1){1.2}}
\put(2.8,0.5){\line(0,1){1.2}}
\put(0.3,0.5){\line(1,1){1.1}}
\put(1.6,0.5){\line(1,1){1.1}}
\put(0.3,1.7){\line(2,-1){2.4}}
\put(-2,2){$A:$}
\put(-2,0){$B:$}
\put(0,0){$x_2^6$}
\put(1.2,0){$x_1^1$}
\put(2.4,0){$x_0^3$}
\put(0,2){$x_0^4$}
\put(1.2,2){$x_2^8$}
\put(2.4,2){$x_0^5$}
\end{picture}  
\begin{picture}(5,4)
\thicklines
\put(0.2,0.5){\line(0,1){1.2}}
\put(1.5,0.5){\line(0,1){1.2}}
\put(2.8,0.5){\line(0,1){1.2}}
\put(0.3,0.5){\line(1,1){1.1}}
\put(1.6,0.5){\line(1,1){1.1}}
\put(0.3,1.7){\line(2,-1){2.4}}
\put(0,0){$x_1^6$}
\put(1.2,0){$x_1^2$}
\put(2.4,0){$x_0^1$}
\put(0,2){$x_1^4$}
\put(1.2,2){$x_0^7$}
\put(2.4,2){$x_1^8$}
\end{picture}  
\begin{picture}(5,4)
\thicklines
\put(0.2,0.5){\line(0,1){1.2}}
\put(1.5,0.5){\line(0,1){1.2}}
\put(2.8,0.5){\line(0,1){1.2}}
\put(0.3,0.5){\line(1,1){1.1}}
\put(1.6,0.5){\line(1,1){1.1}}
\put(0.3,1.7){\line(2,-1){2.4}}
\put(0,0){$x_0^6$}
\put(1.2,0){$x_2^3$}
\put(2.4,0){$x_0^2$}
\put(0,2){$x_2^4$}
\put(1.2,2){$x_1^5$}
\put(2.4,2){$x_2^7$}
\end{picture}  
\end{equation}
\begin{equation}
\begin{picture}(5,4)
\thicklines
\put(0.2,0.5){\line(0,1){1.2}}
\put(1.5,0.5){\line(0,1){1.2}}
\put(2.8,0.5){\line(0,1){1.2}}
\put(0.3,0.5){\line(1,1){1.1}}
\put(1.6,0.5){\line(1,1){1.1}}
\put(0.3,1.7){\line(2,-1){2.4}}
\put(-2,2){$A:$}
\put(-2,0){$B:$}
\put(0,0){$x_2^7$}
\put(1.2,0){$x_1^5$}
\put(2.4,0){$x_2^4$}
\put(0,2){$x_0^2$}
\put(1.2,2){$x_2^3$}
\put(2.4,2){$x_0^6$}
\end{picture}  
\end{equation}
\begin{equation}
\begin{picture}(5,4)
\thicklines
\put(0.2,0.5){\line(0,1){1.2}}
\put(1.5,0.5){\line(0,1){1.2}}
\put(2.8,0.5){\line(0,1){1.2}}
\put(0.3,0.5){\line(1,1){1.1}}
\put(1.6,0.5){\line(1,1){1.1}}
\put(0.3,1.7){\line(2,-1){2.4}}
\put(-2,2){$A:$}
\put(-2,0){$B:$}
\put(0,0){$x_1^3$}
\put(1.2,0){$x_2^1$}
\put(2.4,0){$x_2^2$}
\put(0,2){$x_2^5$}
\put(1.2,2){$x_0^8$}
\put(2.4,2){$x_1^7$}
\end{picture}  
\end{equation}
Each connection means that the corresponding pair appears in the sum (\ref{c3}); there are 8 cycles, each containing 6 connections, which gives 48 terms (two orbits, each containing 24 terms).

Let us assume that for some configuration
\begin{equation}
\sigma:\left.\begin{array}{c}
A=\naw{\ldots x_\alpha^i\ldots}\\
B=\naw{\ldots x_{\alpha'}^{i'}\ldots}
\end{array}\right.\label{bb}
\end{equation}
 the coefficient $c(\sigma)=16$. This means that there are 16 connections between upper and lower rows in (\ref{bb}). Now, any $x$ from one row can be connected to at most two $x$'s of the second row.
This implies that any element of the upper row is connected to exactly two elements of the lower one and reverse. Therefore, the broken line consisting of the connections is a disjoint sum of cycles which is impossible because the number of its vertices (\ref{d1}) is not divisible by 6. Now let us assume that $c(\sigma)=15$. Then in any row of (\ref{bb}) there are 7 vertices with two connections and 1 vertex with 1 connection. If the $"$single$"$ vertices are directly connected, the rest of connections form a broken line with 14 $"$double$"$ vertices which is again impossible as 14 is also not divisible by 6. It is also not difficult to see that the single vertex cannot be directly connected to double one. On the other hand, it is quite easy  to find a configuration $\sigma$ such that $c(\sigma)=14$. For example, this is the case for 
\begin{equation}
\sigma:\left.\begin{array}{c}
A : \quad x_2^1\, x_2^2\, x_1^3\, x_2^4 \, x_1^5 \, x_0^6 \, x_2^7 \, x_0^8\\
B:\quad x_2^1\, x_0^2\, x_2^3\, x_2^4\, x_2^5 \, x_0^6\, x_1^7\, x_0^8
\end{array}\right.\label{ab1}.
\end{equation}
The above reasoning  is supported by explicit computer calculations. Their results are summarized in Table 1\footnote{I am grateful to Dr. S. Sobieski for the help in performing the computer calculations.}.\\
\begin{table}[!ht]
\caption{The number of terms on the right-hand side of eq. (\ref{ab}) with the given value of the coefficient $c(\sigma)$}
\center
\begin{tabular}{|c|r|}
\hline
$c(\sigma)$ & No. of terms in eq. (\ref{ab}) \\
\hline
1 & 327 600\\
2 & 1 494 180\\
3 & 4 141 080\\
4 & 7 754 904\\
5 & 9 832 752\\
6 & 9 010 692\\
7 & 5 984 856\\
8 & 2 966 364\\
9 & 1 094 688\\
10 & 314 712\\
11 & 72 720\\
12 & 12 410\\
13 & 1 584\\
14 & 144\\
15 & 0\\
16 & 0\\
\hline
\end{tabular}
\end{table}
We have shown that the Bell inequality is violated (although only slightly) with our choice of orbits defining the quantum states.

\section{Interpretation in terms of nonlocal game}
Our findings can be rephrazed in the standard way as a nonlocal game (cf. Ref. \cite{ugur1}). There are two players, Alice and Bob, which receive three bits, $s=1,2,\ldots,8$ (Alice) and $t=1,2,\ldots,8$ (Bob) from an arbitrator and then each transmits to the arbitrator a $0$, a $1$ or a $2$. The winning conditions are deduced from the orbit structure, eqs. (\ref{c1}) and (\ref{c2}), and are presented in Table 2.

\begin{table}[!htb]
\caption{Winning values for nonlocal game defined by two orbits of $S_4$}
\center
\small
\begin{tabular}{|l|l|}
\hline
s, t & Alice, Bob \\
\hline
14 & 01\\
15 & 10\\
17 & 21\\
18 & 01, 12, 20\\
24 & 02\\
25 & 22\\
27 & 02, 10, 21\\
28 & 11\\
34 & 00\\
35 & 00, 12, 21\\
37 & 22\\
38 & 10\\
41 & 10\\
42 & 20\\
43 & 00\\
46 & 02, 11, 20\\
\hline
\end{tabular}
\begin{tabular}{|l|l|}
\hline
s, t & Alice, Bob \\
\hline
51 & 01\\
52 & 22\\
53 & 00, 12, 21\\
56 & 10\\
64 & 02, 11, 20\\
65 & 01\\
67 & 10\\
68 & 22\\
71 & 12\\
72 & 01, 12, 20\\
73 & 22\\
76 & 01\\
81 & 02, 10, 21\\
82 & 11\\
83 & 01\\
86 & 22\\
\hline
\end{tabular}
\end{table}
We assume that all 64 values of $(s,t)$ are equally likely. Following the reasoning of Ref. \cite{ugur1} it is now easy to see that the maximal probability of winning the game is, in the classical case, $\frac{14}{64}=\frac{7}{32}$. Each classical deterministic strategy is specified by two functions $f_A(s)$, $f_B(t)$ taking their values, in the set $\poisson{0,1,2}$. One of the optimal strategies can be read off from the eq. (\ref{ab1}). It is summarized in Table 3.
\begin{table}[!htb]
\caption{Optimal classical strategy in $S_4$ game}
\center
\begin{tabular}{|c|c|c|c|c|c|c|c|c|}
\hline
$s$ or $t$ & 1 & 2 & 3 & 4 & 5 & 6 & 7 & 8\\
\hline
$f_A(s)$ & 2 & 2 & 1 & 2 & 1 & 0 & 2 & 0\\
\hline
$f_B(t)$ & 2 & 0 & 2 & 2 & 2 & 0 & 1 & 0\\
\hline 
\end{tabular}
\end{table}
In the quantum strategy Alice and Bob share the state $\ket {\psi}$ spanning the trivial component of $D\otimes D$ in the decomposition into irreducible pieces. They make $a_s$ and $b_t$ measurements according to the information  received from the arbitrator and send the results back to him/her. The probability of winning is now $\frac{14.036}{64}=\frac{7.018}{32}$ which slightly outperforms the classical efficiency.

\section{Conclusions}
In the present paper we gave, following the ideas presented in Refs. \cite{ugur} and \cite{ugur1}, further example of the use of group theoretical methods to analyse the violation of Bell inequalities on the quantum level. Our example uses the symmetric group $S_4$ viewed as the symmetry of regular tetrahedron. This allows us to construct the relevant states and observables on the level of Euclidean geometry in three dimensions. The resulting sum of probabilities entering the Bell inequality exhibits the symmetry of tetrahedron. The degree of violation of the Bell inequality on quantum level can be then established using well known methods of group representations theory.

Our example is based on specific choice of two orbits. The question arises how do the results (i.e. the degree of violation of Bell inequality) depend on the number of orbits considered and their structure; note that although the orbits are for both parties always the same, we "shift" them with respect to each other (by choosing $\ket{\varphi}$ and $\ket{\psi}$ as different members of one party orbit) when constructing the subsequent orbits for the whole system.

In general, the form of Bell inequality and degree of its violation depends on: (i) the choice of symmetry group $G$; (ii) the choice of irreducible representation(s) of $G$ and (iii) the choice of orbit(s). We hope to give more exhaustive analysis elsewhere. Another open problem is whether the group theoretical construction gives the maximal value of violation of Bell inequality. We do not expect this is the case because the quantum system we are considering is special: it exhibits the $S_4$ symmetry. The left hand side of Bell inequality (\ref{w}) carries $S_4$ symmetry acting by permuting the probabilities entering the sum (because the probabilities emerge from the orbit structure). However, it is valid for any set of sixteen observables, each attaining three values. Therefore, in looking for Cirel'son bound one should consider all appropriate quantum systems, not only those carrying $S_4$ symmetry.

Our construction can be extended to all Platonic solids. In fact, the symmetry of the cube and octahedron is the direct product of $S_4$ and space inversion while that of dodecahedron and icosahedron - the direct product of alternating group $A_5$ and space inversion. The Platonic solids are the degenerate orbits in appriopriate representations of their symmetry groups and, again, we can use this fact to construct, by means of threedimensional Euclidean geometry, the relevant orbits defining the relevant observables.

The Bell inequalities considered here are the inequalities of the form $S\leq C$ where $S$ is a certain sum of joint probabilities (cf. eq. (\ref {c3})) while $C$ is some constant. As explained in the Introduction on the example of CHSH inequalities such an inequality can be converted into the standard form involving certain correlation functions. To this end one should express the relevant probabilities in terms of correlation functions, more or less complicated (cf. also Ref. \cite{fine}, \cite{fine1}).
Let us note that the correlation functions one has to use depend on the dimension of the space of states. In our case the observables $a_i$ (or $b_i$) obey the third order polynomial equation. Therefore, the independent correlation functions involve the products $a_i^{\sigma_i}b_j^{\rho_j}$, $0\leq\sigma_i\leq 2$, $0\leq\rho_j\leq 2$ which allow to express the relevant probabilities in terms of them.

Finally, we conjecture that the similar construction, exhibiting $S_n$ symmetry, is possible for arbitrary $n$. $S_n$ is the symmetry group of the simplest regular $n-1$  -dimensional polyhedron consisting of $n$ vertices and $\binom{n}{2}$ edges. It acts by permuting the vertices through reflections and rotations and their combinations. This is the so-called standard representation. Again, one should look for the orbit consisting of $n\naw{n-2}!$ \,\,$n-1$-tuples of mutually orthogonal vectors and proceed as in $S_3$ and $S_4$ cases.   

\subsection*{Acknowledgements}
I would like to thank Prof. Piotr Kosi\'nski  for helpful discussion and useful remarks. I gratefully acknowledge interesting discussions with Dr K. Smoli\'nski and Prof. Z. Walczak. Special thanks are to Dr. \'S. Sobieski for invaluable help in numerical calculations. 
 This research is supported by the NCN Grant no. DEC-2012/05/D/ST2/00754.

\begin{appendices}
\section{}
Below we present the explicit form of the vectors $\ket{x_{\alpha}^i}$, $\alpha=0,1,2$, $i=1,\ldots,8$ entering the orbit constructed in Sec. II. They read (cf. eq. (\ref{b5})):\\
\begin{footnotesize}
$a_1:\quad \naw{x_0^1,x_1^1,x_2^1}$
\begin{equation}
\ket{x_0^1}=\left[\begin{array}{c}
\frac{\sqrt{3}}{3}\\
\frac{\sqrt{3}}{3}\\
-\frac{\sqrt{3}}{3}
\end{array}\right],
\quad \ket{x_1^1}=\left[\begin{array}{c}
\frac{\sqrt{3}}{3}\\
\frac{1}{2}\naw{1-\frac{\sqrt{3}}{3}}\\
\frac{1}{2}\naw{1+\frac{\sqrt{3}}{3}}\\
\end{array}\right],\quad 
\ket{x_2^1}=\left[\begin{array}{c}
\frac{\sqrt{3}}{3}\\
-\frac{1}{2}\naw{1+\frac{\sqrt{3}}{3}}\\
-\frac{1}{2}\naw{1-\frac{\sqrt{3}}{3}}\\
\end{array}\right]
\end{equation}
$a_2:\quad \naw{x_0^2,x_1^2,x_2^2}$
\begin{equation}
\ket{x_0^2}=\left[\begin{array}{c}
\frac{1}{9}\naw{-3\sqrt{2}-\sqrt{3}-\sqrt{6}}\\
\frac{1}{18}\naw{-3+5\sqrt{3}-2\sqrt{6}}\\
\frac{1}{6}\naw{-1-2\sqrt{2}+\sqrt{3}}\\
\end{array}\right],\,
 \ket{x_1^2}=\left[\begin{array}{c}
\frac{1}{9}\naw{-\sqrt{3}+2\sqrt{6}}\\
\frac{1}{18}\naw{9-\sqrt{3}-2\sqrt{6}}\\
-\frac{1}{6}\naw{1+2\sqrt{2}+\sqrt{3}}\\
\end{array}\right],\,
\ket{x_2^2}=\left[\begin{array}{c}
\frac{1}{9}\naw{3\sqrt{2}-\sqrt{3}-\sqrt{6}}\\
-\frac{1}{9}\naw{3+2\sqrt{3}+\sqrt{6}}\\
\frac{1}{3}\naw{1-\sqrt{2}}\\
\end{array}\right],
\end{equation}
$a_3:\quad \naw{x_0^3,x_1^3,x_2^3}$
\begin{equation}
\ket{x_0^3}=\left[\begin{array}{c}
\frac{1}{9}\naw{3\sqrt{2}-\sqrt{3}-\sqrt{6}}\\
\frac{1}{18}\naw{3+5\sqrt{3}-2\sqrt{6}}\\
\frac{1}{6}\naw{1+2\sqrt{2}+\sqrt{3}}\\
\end{array}\right],\,
 \ket{x_1^3}=\left[\begin{array}{c}
\frac{1}{9}\naw{-\sqrt{3}+2\sqrt{6}}\\
-\frac{1}{18}\naw{9+\sqrt{3}+2\sqrt{6}}\\
\frac{1}{6}\naw{1+2\sqrt{2}-\sqrt{3}}\\
\end{array}\right],\,
\ket{x_2^3}=\left[\begin{array}{c}
-\frac{1}{9}\naw{3\sqrt{2}+\sqrt{3}+\sqrt{6}}\\
\frac{1}{9}\naw{3-2\sqrt{3}-\sqrt{6}}\\
\frac{1}{3}\naw{-1+\sqrt{2}}\\
\end{array}\right],
\end{equation}
$a_4:\quad \naw{x_0^4,x_1^4,x_2^4}$
\begin{equation}
\ket{x_0^4}=\left[\begin{array}{c}
\frac{1}{9}\naw{3\sqrt{2}-\sqrt{3}-\sqrt{6}}\\
\frac{1}{18}\naw{3-\sqrt{3}+4\sqrt{6}}\\
\frac{1}{6}\naw{3+\sqrt{3}}\\
\end{array}\right],\,
 \ket{x_1^4}=\left[\begin{array}{c}
\frac{1}{9}\naw{-\sqrt{3}+2\sqrt{6}}\\
\frac{1}{9}\naw{\sqrt{3}+2\sqrt{6}}\\
-\frac{\sqrt{3}}{3}
\end{array}\right],\,
\ket{x_2^4}=\left[\begin{array}{c}
-\frac{1}{9}\naw{3\sqrt{2}+\sqrt{3}+\sqrt{6}}\\
\frac{1}{18}\naw{-3-\sqrt{3}+4\sqrt{6}}\\
\frac{1}{6}\naw{-3+\sqrt{3}}\\
\end{array}\right],
\end{equation}
$a_5:\quad \naw{x_0^5,x_1^5,x_2^5}$
\begin{equation}
\ket{x_0^5}=\left[\begin{array}{c}
\frac{1}{9}\naw{-\sqrt{3}+2\sqrt{6}}\\
\frac{1}{18}\naw{9-\sqrt{3}-2\sqrt{6}}\\
\frac{1}{6}\naw{1+2\sqrt{2}+\sqrt{3}}\\
\end{array}\right],\,
 \ket{x_1^5}=\left[\begin{array}{c}
-\frac{1}{9}\naw{3\sqrt{2}+\sqrt{3}+\sqrt{6}}\\
\frac{1}{18}\naw{-3+5\sqrt{3}-2\sqrt{6}}\\
\frac{1}{6}\naw{1+2\sqrt{2}-\sqrt{3}}\\
\end{array}\right],\,
\ket{x_2^5}=\left[\begin{array}{c}
\frac{1}{9}\naw{3\sqrt{2}-\sqrt{3}-\sqrt{6}}\\
-\frac{1}{9}\naw{3+2\sqrt{3}+\sqrt{6}}\\
\frac{1}{3}\naw{-1+\sqrt{2}}\\
\end{array}\right],
\end{equation}
$a_6:\quad \naw{x_0^6,x_1^6,x_2^6}$
\begin{equation}
\ket{x_0^6}=\left[\begin{array}{c}
-\frac{1}{9}\naw{3\sqrt{2}+\sqrt{3}+\sqrt{6}}\\
\frac{1}{18}\naw{-3-\sqrt{3}+4\sqrt{6}}\\
\frac{1}{6}\naw{3-\sqrt{3}}
\end{array}\right],\,
 \ket{x_1^6}=\left[\begin{array}{c}
\frac{1}{9}\naw{3\sqrt{2}-\sqrt{3}-\sqrt{6}}\\
\frac{1}{18}\naw{3-\sqrt{3}+4\sqrt{6}}\\
-\frac{1}{6}\naw{3+\sqrt{3}}
\end{array}\right],\,
\ket{x_2^6}=\left[\begin{array}{c}
\frac{1}{9}\naw{-\sqrt{3}+2\sqrt{6}}\\
\frac{1}{9}\naw{\sqrt{3}+2\sqrt{6}}\\
\frac{\sqrt{3}}{3}
\end{array}\right],
\end{equation}
$a_7:\quad \naw{x_0^7,x_1^7,x_2^7}$
\begin{equation}
\ket{x_0^7}=\left[\begin{array}{c}
\frac{1}{9}\naw{3\sqrt{2}-\sqrt{3}-\sqrt{6}}\\
\frac{1}{18}\naw{3+5\sqrt{3}-2\sqrt{6}}\\
-\frac{1}{6}\naw{1+2\sqrt{2}+\sqrt{3}}
\end{array}\right],\,
 \ket{x_1^7}=\left[\begin{array}{c}
\frac{1}{9}\naw{-\sqrt{3}+2\sqrt{6}}\\
-\frac{1}{18}\naw{9+\sqrt{3}+2\sqrt{6}}\\
\frac{1}{6}\naw{-1-2\sqrt{2}+\sqrt{3}}\\
\end{array}\right],\,
\ket{x_2^7}=\left[\begin{array}{c}
-\frac{1}{9}\naw{3\sqrt{2}+\sqrt{3}+\sqrt{6}}\\
\frac{1}{9}\naw{3-2\sqrt{3}-\sqrt{6}}\\
\frac{1}{3}\naw{1-\sqrt{2}}\\
\end{array}\right],
\end{equation}
$a_8:\quad \naw{x_0^8,x_1^8,x_2^8}$
\begin{equation}
\ket{x_0^8}=\left[\begin{array}{c}
\frac{\sqrt{3}}{3}\\
-\frac{1}{2}\naw{1+\frac{\sqrt{3}}{3}}\\
\frac{1}{2}\naw{1-\frac{\sqrt{3}}{3}}
\end{array}\right],\,
 \ket{x_1^8}=\left[\begin{array}{c}
\frac{\sqrt{3}}{3}\\
\frac{1}{2}\naw{1-\frac{\sqrt{3}}{3}}\\
-\frac{1}{2}\naw{1+\frac{\sqrt{3}}{3}}
\end{array}\right],\,
\ket{x_2^8}=\left[\begin{array}{c}
\frac{\sqrt{3}}{3}\\
\frac{\sqrt{3}}{3}\\
\frac{\sqrt{3}}{3}
\end{array}\right].
\end{equation}
\end{footnotesize}

\end{appendices}


\begin{thebibliography}{99}
\bibitem{bell} J.S. Bell, \emph{Physics} \textbf{1}, 195 (1964)
\bibitem{clauser} J.F. Clauser, M.A. Horne, A. Shimony, R.A. Holt, \emph{Phys. Rev. Lett.} \textbf{23}, 880 (1969)
\bibitem{clauser1} J.F. Clauser, M.A. Horne, \emph{Phys. Rev.} \textbf{ D10}, 526 (1974)
\bibitem{kaszlikowski} D. Kaszlikowski, P. Gnacinski, M. Zukowski, W. Miklaszewski, A. Zeilinger, \emph{Phys. Rev. Lett.} \textbf{85}, 4418 (2000)
\bibitem{werner} R.F. Werner, M.M. Wolf, \emph{Phys. Rev.} \textbf{A64}, 032112 (2001)
\bibitem{collins} D. Collins, N. Gisin, N. Linden, S. Massar, S. Popescu, \emph{Phys. Rev. Lett.} \textbf{88}, 040404 (2002)
\bibitem{son} W. Son, J. lee, M.S. Kim, \emph{Phys. Rev. Lett.} \textbf{96}, 060406
\bibitem{cabello} A. Cabello, S. Severini, A. Winter, \emph{Phys. Rev. Lett.} \textbf{112}, 040401 (2014)
\bibitem{liang} Y-C. Liang, R. Spekkens, H. Wiseman, \emph{Phys. Rep.} \textbf{506}, 1 (2011)
\bibitem{brunner} N. Brunner, D. Cavalcanti, S. Pironio, V. Scami, S. Wehner, \emph{Rev. Mod. Phys.} \textbf{86}, 419 (2014)
\bibitem{ekert} A.K. Ekert, \emph{Phys. Rev. Lett.} \textbf{67}, 661 (1997)
\bibitem{fine} A. Fine, \emph{Phys. Rev. Lett.} \textbf{48}, 291 (1982)
\bibitem{fine1} A. Fine, \emph{Journ. Math. Phys.} \textbf{23}, 1306 (1982)
\bibitem{halliwell} J.J. Halliwell, J.M. Yearsley, \emph{Phys. Rev.} \textbf{A87}, 022114 (2013)
\bibitem{halliwell1} J.J. Halliwell, \emph{Phys. Lett.} \textbf{A378}, 2945 (2014)
\bibitem{ugur} V. Ug\v ur G\H uney, M. Hillery, \emph{Phys. Rev.} \textbf{A90}, 062121 (2014)
\bibitem{ugur1} V. Ug\v ur G\H uney, M. Hillery, \emph{Phys. Rev.} \textbf{A91}, 052110 (2015)
\bibitem{cirelson} B.S. Cirel'son, \emph{Lett. Math. Phys.} \textbf{4}, 93 (1980)

\end{thebibliography}
\end{document}